\documentclass[11pt]{article} 

\usepackage{cancel}
\usepackage{amsmath}

\usepackage[center,font=small,labelfont=bf]{caption}

\DeclareFontFamily{OT1}{pzc}{}
\DeclareFontShape{OT1}{pzc}{m}{it}%
              {<-> s * [0.900] pzcmi7t}{}
\DeclareMathAlphabet{\mathpzc}{OT1}{pzc}%
                                 {m}{it}

\usepackage[utf8]{inputenc} 

\usepackage{geometry} 
\usepackage{graphicx} 

\usepackage{booktabs} 
\usepackage{array} 
\usepackage{paralist} 
\usepackage{verbatim} 
\usepackage{subfig} 

\usepackage{fancyhdr} 
\usepackage{sectsty}
\allsectionsfont{\sffamily\mdseries\upshape} 

\usepackage[nottoc,notlof,notlot]{tocbibind} 
\usepackage[titles,subfigure]{tocloft} 

\numberwithin{equation}{subsection}

\title{Introductory Causal Dynamical Triangulation}
\author{Alex Forcier}
\date{} 

\begin{document}

\begin{titlepage}

\begin{center}

\textsc{\LARGE Introductory Causal Dynamical Triangulation}\\[0.5cm]

\Large{A. Forcier}\\[1cm]

A report submitted in conformity with the requirements\\
for the degree of Master of Science\\
Department of Physics\\
University of Toronto\\[2.5cm]

\end{center}

\begin{abstract}
This report aims to present the main ideas of Regge calculus necessary to understand the basic premise of CDT. Next, the main strategy of the CDT approach is introduced in general terms. The main focus of this report is the 2-D model of CDT. The section on the 2-D model closely follows a single paper (\cite{ambjorn98}). While the 4-D or even 3-D case will behave very differently from the 2-D model, 2-D CDT can be solved exactly, and as such offers a better introductory exposition of CDT's methods. Higher-dimensional CDT requires a lot of computer simulation, and lies outside the scope of this report.

All derivations carried out explicitly are the result of the author's independent work in attempting to find and prove how the results presented were obtained by CDT authors. Because these derivations were made explicit by the author, this paper can act as a guide to those who are new to CDT.\\[2cm]
\end{abstract}

\begin{center}
Copyright \copyright 2011 by Alex Forcier
\end{center}

\end{titlepage}

\newpage

\tableofcontents

\newpage

\section{Introduction}

\subsection*{Physical Motivation for CDT}

One of the main goals of modern physics can be said to be the discovery of a consistent description of physical phenomena at all scales. Causal Dynamical Triangulation (CDT) is one suggested avenue of approach to this goal and thus it is of interest. 

The most attractive aspect of CDT is that it offers an approach to deriving the nature of spacetime from a minimal set of assumptions: the entire model arises from only an initial triangulation of spacetime (this puts it in the non-perturbative family of approaches to quantum gravity). In general, the idea of deriving what is observed from first principles (without having to postulate too much about the fundamental nature and structure of space and time) is an attractive one, and makes any model such as CDT that has shown initial promising results worth studying.

\subsection*{Motivation for Studying CDT}

One of the main reasons for studying CDT this year is the interesting physical motivation presented above. Another reason for choosing CDT as a topic of research is that there is surprisingly little commonly known about it and there are apparently not a large amount of people working on it, despite it purportedly being a promising approach to studying how quantum mechanics and gravity may be combined. The aim was to understand why this may be and to understand the main ideas of the model.

\subsection*{Overview and Purpose of This Report}

The goal of this MSc report and the research that shaped it is to gain a basic understanding of the form and function of CDT.

The basic idea of CDT is to discretize spacetime by approximating it with a piecewise linear manifold, a process called triangulation. The building blocks of the new manifold are simplices, which are an arbitrary-dimensional generalization of the notion of a triangle or tetrahedron. One then constructs the path integral describing the evolution of spacetime, which can eventually be used to obtain the physical predictions of the model.

This report aims to present the main ideas of Regge calculus necessary to understand the basic premise of CDT. Next, the main strategy of the CDT approach is introduced in general terms. The main focus of this report is the 2-D model of CDT. The section on the 2-D model closely follows a single paper (\cite{ambjorn98}). While the 4-D or even 3-D case will behave very differently from the 2-D model, 2-D CDT can be solved exactly, and as such offers a better introductory exposition of CDT's methods. Higher-dimensional CDT requires a lot of computer simulation, and lies outside the scope of this report.

All derivations carried out explicitly are the result of the author's independent work in attempting to find and prove how the results presented were obtained by CDT authors. Because these derivations were made explicit by the author, this paper can act as a guide to those who are new to CDT.

\newpage

\section{Regge Calculus}

Before delving into CDT proper, an overview of Regge calculus is necessary in order to provide some of the tools required for CDT. The sources used for most of this section were \cite{mitchell} and \cite{parandis}.

Regge calculus is a discretisation of general relativity in which there are no fields, just a triangulation of spacetime. There are two main motivations for using Regge calculus techniques:

\begin{enumerate}
\item It offers a way of working in general relativity without using symmetries. Ordinarily, significant assumptions of symmetry are required to reduce the complexity of relevant equations to a manageable form.
\item It offers an approach to discretising GR, opening new avenues of approach in the search for a successful theory of quantum gravity.
\end{enumerate}

In the regime of Regge calculus, one considers spacetime to be made up flat (Minkowskian) ``simplices" joined face to face, edge to edge and vertex to vertex. A simplex is a generalization of a triangle (the 2-D simplex) to arbitrary dimension - the 3-simplex is a tetrahedron, a 1-simplex is a line, and a 0-simplex is a point). In this way, a smooth manifold can be approximated arbitrarily closely by joining any number of sufficiently small simplices.

Understanding Regge calculus amounts to studying how triangulated spacetime changes the meaning and measure of curvature, and how that changes the form of related equations in general relativity. The form of objects like the metric tensor under Regge calculus ends up being much simpler: all the information required to know the geometry is given by the lengths of the edges of the simplices involved.

The simplest way to illustrate the curvature of triangulated space is to analyse the 2-D case. Consider using a triangle with geodesic edges to probe the geometry of a 2-D smooth manifold. If the geometry now enclosed inside the triangle is non-Euclidean, the sum of the internal angles of the triangle will deviate from $\pi$. The measure of this deviation is the Gaussian integral curvature $\epsilon_t$, defined for a triangle $t$ with internal angles $\alpha$, $\beta$ and $\gamma$:

\begin{equation}
\epsilon_t = \alpha + \beta + \gamma - \pi.
\end{equation}

We also need the definition of the local Gaussian curvature K at a point P:

\begin{equation}
K(P) = \lim_{A_t \to 0} \frac{\epsilon_t}{A_t},
\end{equation}
where $A_t$ is the triangle's area. Alternatively:

\begin{equation}
\epsilon_t = \int_{t} K(P) d A.
\end{equation}

\begin{figure}
\begin{center}
\includegraphics[scale=0.5]{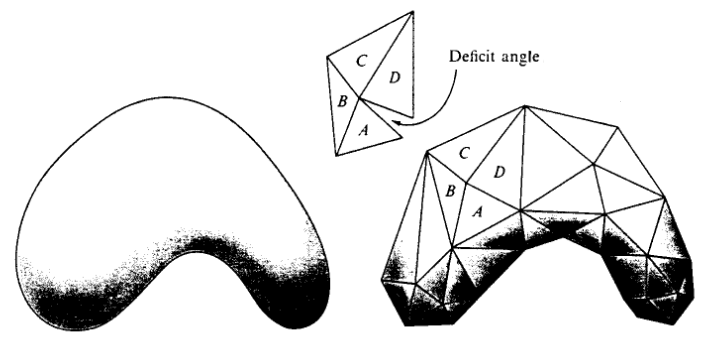} 
\end{center}
\caption{Triangulation of a 2-D manifold \\
\cite{misner73}}
\end{figure}

Let us now consider instead a smooth 2-D manifold whose surface we have triangulated - that is to say, we have approximated the original smooth surface by connecting flat triangles along their edges (see Figure 1). Using the same triangle test mentioned earlier to probe the geometry of what is now a polyhedral approximation to the earlier smooth manifold, we are led to three possible cases:

\begin{enumerate}[I]
\item The triangle is placed entirely within one of the triangles composing the triangulation.
\item The triangle crosses an edge where two triangles connect, but encloses no vertices.
\item The triangle encloses at least one vertex.
\end{enumerate}

Since each face of the polyhedron is a flat triangle, the triangle in the first case encloses no curvature. The second case corresponds to taking a flat triangle and folding it - again no curvature is introduced. That leaves the third case: the only possibility of the test triangle enclosing curvature is if it encloses vertices. Thus all curvature is concentrated at the vertices: $K(P) = 0$ if $P$ is not a vertex (note, however, that the converse is not necessarily true: $K(P)=0$ can still occur if $P$ is a vertex, as triangles may join at a vertex without generating any curvature).

Of course any integral curvature is not a property of the test triangle but of the vertex; we will instead call it the deficiency of the vertex $V$, labelled $\epsilon_V$. If a test triangle t encloses more than one vertex, $\epsilon_t$ is the sum of the deficiencies of the enclosed vertices.

An alternate way of measuring the deficiency of a vertex is by computing the sum of the internal angles of faces that meet at the vertex - they add up to $2 \pi - \epsilon_V$. This means that the same curvature information obtained by means of the geodesic test triangle can be obtained from the triangulation itself.

The geodesic test triangle is a useful illustrative tool in 2 dimensions, but for higher dimensional spaces it is much more useful to employ the notion of parallel transporting a vector around a closed loop. The angle $\sigma(a)$ between a vector and the same vector after parallel transport around a closed loop $a$ is given by:

\begin{equation}
\sigma(a) = \int_a K d A,
\end{equation}
where $A$ is the area of the loop $a$.

In terms of our 2-D considerations above, the vector will be unchanged after parallel transport unless the loop contains a vertex, and upon comparing equation (2.0.4) to (2.0.3) we see that the angle of rotation of the parallel transported vector is exactly the deficit angle of the vertex (or the sum of the deficit angles of the vertices).

The purpose of introducing parallel transport, however, was for higher-dimensional space considerations. In 3-D, our triangulation is formed now by 3-simplices (tetrahedra) connected along their flat triangular faces (2-simplices). Parallel transport of a vector around a loop confined to a single simplex will not change the vector; nor will the vector be changed for a loop that passes through a second tetrahedron but returns through the same face, enclosing no edges. The only way the vector can be affected is if the loop encloses one of the 1-simplex edges (`bones', in Regge's terminology) shared by faces of multiple 3-simplices. The deficiency of the edge is found by computing the sum of the angles between the 2-simplices meeting at the 1-simplex: the sum will be $2 \pi - \epsilon$, where $\epsilon$ is the deficiency of the edge.

One can see how a this regime can be extrapolated to any dimension: a simplicial spacetime is constructed by connecting d-simplices along flat (d-1)-sub-simplices. The curvature is concentrated to (d-2)-sub-simplices (bones) and is computed as a deficiency angle by summing up the angles between the (d-1)-simplices that meet at the bone. 

\subsection*{Regge Riemann Curvature Tensor}

The next step is to translate some general relativity concepts into Regge calculus terms. The first is the Riemann curvature tensor, which will be used to obtain the Regge action.

Each bone is a (d-2)-sub-simplex; therefore there exists a 2-D plane perpendicular to it. For illustrative purposes we will confine ourselves to three dimensions. Following \cite{mitchell}, we consider a bundle of parallel bones (now lines) with a high density of bones per unit area $\rho$, each with the same deficit angle $\epsilon$.

$\vec{U}$ is a unit vector parallel to the bones, and $\vec{A}$ is parallel transported around a small loop of area $\Sigma$ with unit normal $\vec{n}$ and area $\vec{\Sigma}=\vec{n} \Sigma$. Decompose $\vec{A}$ into $\vec{A}^{\parallel}$ parallel to $\vec{U}$ and $\vec{A}^{\perp}$ lying in the plane perpendicular to $\vec{U}$. Under parallel transport around $\Sigma$ only $\vec{A}^{\perp}$ is affected, rotated by an angle $\sigma=N\epsilon$, where N is the number of bones piercing the loop. In this manner, $\vec{A}$ is rotated about $\vec{U}$ by an angle $\sigma$.

\begin{figure}
\begin{center}
\includegraphics[scale=0.5]{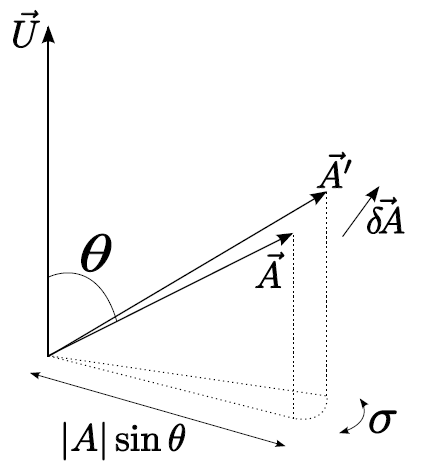} 
\end{center}
\caption{The vector $\vec{A}$ is rotated about $\vec{U}$ by an angle $\sigma$. \\
\cite{mitchell}}
\end{figure}

From Figure 2 ($\theta$ is the angle from $\vec{U}$ to $\vec{A}$), $|\delta A|=\sigma |A| \sin\theta=\sigma |U| |a| \sin\theta$ (because $\vec{U}$ is a unit vector). Thus:

\begin{equation}
\vec{\delta A} = \sigma \left( \vec{U} \times \vec{A} \right).
\end{equation}

Now $\sigma=N \epsilon$ and the number of bones crossing $\Sigma$ is given by $N=\rho \vec{U} \cdot \vec{\Sigma}$, we can write the above as:

\[
\vec{\delta A} = \rho \epsilon\left( \vec{U} \cdot \vec{\Sigma} \right)  \left( \vec{U} \times \vec{A} \right)
\]
or, in component form:

\begin{equation}
\delta A_{\mu} = \rho \epsilon \epsilon_{\mu\nu\sigma} U^\nu A^\sigma U^\gamma \Sigma_\gamma,
\end{equation}
where $\epsilon_{\mu\nu\sigma}$ is the anti-symmetric tensor.

When a vector $\vec{A}$ is parallel transported around an infinitesimal parallelogram with sides $\vec{d x}$ and $\vec{d y}$ then $\vec{\delta A}$ is given by:

\begin{equation}
\delta A_\mu = R_{\mu\alpha\beta}^{\sigma} A_{\sigma} d x^\alpha d y^\beta = R_{\mu\alpha\beta}^{\sigma} A_{\sigma} \frac{1}{2} \left( d x^\alpha d y^\beta - d x^\beta d y^\alpha \right),
\end{equation}
where the symmetry property of the Riemann tensor was used. The loop area $\vec{\Sigma}$ of the parallelogram is given by:

\[
\vec{\Sigma} = \vec{d x} \times \vec{d y}
\]
or, in component form:

\begin{equation}
\Sigma_\mu = \epsilon_{\mu\nu\rho} d x^\nu d y^\rho.
\end{equation}

Using the identity of the anti-symmetric tensor $\epsilon^{\alpha\beta\mu} \epsilon_{\mu\nu\rho} = \delta_{\nu}^{\alpha} \delta_{\rho}^{\beta} - \delta_{\nu}^{\beta} \delta_{\rho}^{\alpha}$,

\[
\epsilon^{\alpha\beta\mu} \Sigma_\mu = d x^\alpha d y^\beta - d x^\beta d y^\alpha.
\]

Substituting into (2.0.7):

\begin{equation}
\delta A_\mu = \frac{1}{2} R_{\mu\alpha\beta}^{\sigma} \epsilon^{\alpha\beta\gamma} \Sigma_\gamma A_{\sigma} .
\end{equation}

Comparing this expression to (2.0.6), we find:

\[
\begin{split}
\frac{1}{2} R_{\sigma\mu\alpha\beta} \epsilon^{\alpha\beta\gamma} \Sigma_\gamma A^{\sigma} &= \rho \epsilon \epsilon_{\mu\nu\sigma} U^\nu A^\sigma U^\gamma \Sigma_\gamma \\
\Rightarrow R_{\sigma\mu\alpha\beta} \epsilon^{\alpha\beta\gamma} &= 2 \rho \epsilon \epsilon_{\mu\nu\sigma} U^\nu U^\gamma.
\end{split}
\]

Multiplying both sides by $\epsilon_{\gamma\rho\eta}$ and using again the identity $\epsilon^{\alpha\beta\gamma} \epsilon_{\gamma\rho\eta} = \delta_{\rho}^{\alpha} \delta_{\eta}^{\beta} - \delta_{\eta}^{\alpha} \delta_{\rho}^{\beta}$:

\[
R_{\sigma\mu\rho\eta} - R_{\sigma\mu\eta\rho} = 2 \rho \epsilon (-U_{\mu\sigma}) U_{\rho\eta},
\]
where $U_{\rho\eta}=\epsilon_{\rho\eta\gamma} U^\gamma$.

Therefore we have:

\begin{equation}
R_{\sigma\mu\rho\eta} = \rho \epsilon U_{\sigma\mu} U_{\rho\eta}.
\end{equation}

\subsection*{The Regge Action}

To obtain the Regge action, we begin with the regular Einstein-Hilbert action:

\begin{equation}
S_{E H}[g] = \frac{1}{16 \pi G} \int d^d x \sqrt{-g} (R-2\Lambda).
\end{equation}

Let us first consider only the $R$-dependent part of the action. In our simplicial spacetimes the curvature is restricted to the bones; therefore, we expect that the equivalent of scalar curvature $R$ under the Regge regime will vanish everywhere except on the bones $b_n$. Thus we expect the $R$-dependent part of the action to take the form:

\[
S_{Regge_R} = \sum_n F(b_n),
\]
where the sum is over all bones and the function $F$ is related to curvature and must be the same for all bones. We can expect $F(b_n)$ to be proportional to the volume of the bone $k_n$ since the bones are homogeneous. As was stressed earlier, the only curvature-related variable we have is the deficit angle; we can thus conclude that $F(b_n)$ must have the form:

\[
F(b_n) = k_n f(\epsilon_n).
\]

To determine the function $f$, consider the following argument: any bone $b_n$ can be represented as the superposition of two bones $b_{n\prime}$ and $b_{n\prime\prime}$, of the same shape and area such that $\epsilon_n = \epsilon_{n\prime} + \epsilon_{n\prime\prime}$. This requires:

\[
f(\epsilon_n) = f(\epsilon_{n\prime}) + f(\epsilon_{n\prime\prime}),
\]
which implies $f(\epsilon)=C\epsilon$ for some constant $C$. Our ($R$-dependent) Regge action is then:

\begin{equation}
S_{Regge_R} = C \sum_n \epsilon_n k_n.
\end{equation}

We determine the value of $C$ by comparing this result to the Einstein-Hilbert action at high bone density. Under that constraint the relationship between $R$ and the deficit angle is described by (2.0.10). Using the identity $U_{\mu\sigma} U^{\mu\sigma}=2$, we find:

\[
R = 2 \rho \epsilon.
\]

Substituting this into the $R$-dependent part of the Einstein-Hilbert action:

\[
\frac{1}{16 \pi G} \int d^d x \sqrt{-g} R \rightarrow \frac{1}{8 \pi G} \int d^d x \sqrt{-g} \rho \epsilon.
\]
Recalling that $\rho$ is the density of bones per unit area, corresponding to a 2-D plane perpendicular to a cluster of (d-2)-dimensional bones, we conclude that integrating over all space gives the total volume of bones:

\[
\int d^d x \sqrt{-g} \rho \epsilon = \sum_n k_n \epsilon_n,
\]
where, as before, $k_n$ is the (d-2)-dimensional volume of the bone. Thus,

\[
\frac{1}{16 \pi G} \int d^d x \sqrt{-g} R \rightarrow \frac{1}{8 \pi G} \sum_n \epsilon_n k_n \Rightarrow C=\frac{1}{8 \pi G}
\]
by comparing with (2.0.12). Therefore:

\begin{equation}
S_{Regge_R} = \kappa \sum_n \epsilon_n k_n, \quad \kappa = \frac{1}{8 \pi G}.
\end{equation}

Consider now the term proportional to $\Lambda$ in the Einstein-Hilbert action:

\[
\frac{-2 \Lambda}{16 \pi G} \int d^d x \sqrt{-g} \rightarrow \frac{-2 \Lambda}{16 \pi G} \sum_n V_n,
\]
where we have recognized that the integral gives us the volume of the entire simplicial spacetime, which we can express as a sum over the volume of all $n$-simplices composing the spacetime. Thus:

\begin{equation}
S_{Regge_{\Lambda}} = -\lambda \sum_n V_n, \quad \lambda = \frac{\Lambda}{8 \pi G}.
\end{equation}

Finally, combining (2.0.13) and (2.0.14):

\begin{equation}
S_{Regge} = \sum_n \left( \kappa \epsilon_n k_n - \lambda V_n \right).
\end{equation}

This is the general expression for the Regge action in any simplicial spacetime dimension. Some refinements which will prove useful can be made in the 2-D case, which we will be focusing on later on.

Recalling equation (2.0.3), we can write:

\[
S_{Regge} = \sum_n \left( \kappa \int_t K(P) d A - \lambda V_n \right),
\]
where the $k_n$ term has been dropped because a bone in a 2-D simplicial spacetime is a point, and therefore has no volume content. Employing the discrete Gauss-Bonnet theorem:

\begin{equation}
\sum_k \int_k K(P) d A = 2 \pi \chi,
\end{equation}
where $\chi$ is the Euler characteristic and can be written as $\chi = 2-2 \mathpzc{g}$, where $\mathpzc{g}$ is the genus (the number of holes) of the surface.

Thus:

\[
S_{Regge_{2 D}} = \kappa \, 2 \pi \chi - \sum_n \lambda V_n
\]

\begin{equation}
\Rightarrow S_{Regge_{2 D}} = \kappa \, 2 \pi \chi - N(T) \lambda A.
\end{equation}

In the final equation we have eliminated the sum by expressing $\sum_n V_n = N(T) A$, where $N(T)$ is the number of triangles and $A$ the area of each triangle. This is because all the triangles have the same area (in two dimensions the ``volume" of the simplices becomes the area of the triangles).

\newpage

\section{CDT Overview}

The aim of this section is to provide a general overview of the goals and methods of CDT, following mostly \cite{ambjorn06}.

In the quest for understanding of the structure of spacetime at smaller and smaller scales, one important conclusion we can draw is based on two premises:

\begin{enumerate}[I]
\item Due to the Heisenberg uncertainty principle, probing at shorter distances introduces larger energy-momentum fluctuations.
\item According to general relativity, the presence of energy fluctuations will deform the geometry of the spacetime, imparting curvature.
\end{enumerate}

The combination of the two leads us to conclude that the structure of spacetime at the Planck scale must be highly curved and dynamical.

Research into quantum gravity can be divided into two general categories:

\begin{enumerate}[(a)]
\item non-perturbative: quantize the gravitational degrees of freedom, without adding any additional structure.
\item string-theoretic: the quantization of gravity appears as a by-product of a unified higher-dimensional, supersymmetric theory.
\end{enumerate}

CDT belongs in the first category. As will be summarized in this section, CDT has produced a few appealing tangible results so far: mainly, there is evidence that the theory has a good classical limit and the theory has provided first indications of what description it offers of quantum structure at Planck scale.

The two main tools required to do CDT are path integrals and Regge calculus. The general form for a path integral describing the sum over the virtual paths taken by a particle between initial and final points ${\bf x_i}$ and ${\bf x_f}$ is:

\begin{equation*}
G({\bf x_i},{\bf x_f};t) = \sum_{{\rm paths}:{\bf x_i} \rightarrow {\bf x_f}} e^{i S^{{\rm part}}[x(\tau)]},
\tag{3.0.1}
\end{equation*}
where $S^{{\rm part}}$ describes the action associated with the particle.

For gravity, the path integral would be a superposition of all virtual "paths" the universe can follow as time unfolds. In this case the paths are the different configurations of the metric field variables $g_{\mu\nu}(x)$ (accounting only for gravitational degrees of freedom; which is to say, ignoring matter fields). Thus we express the path integral for gravity in the generic form:

\begin{equation*}
G({\bf g_i},{\bf g_f};t) = \sum_{{\rm spacetimes}:{\bf g_i} \rightarrow {\bf g_f}} e^{i S^{{\rm grav}}[g_{\mu\nu}({\bf x},\tau)]},
\tag{3.0.2}
\end{equation*}
where $S^{{\rm grav}}$ denotes the gravitational action associated with a metric $g_{\mu\nu}$ with initial and final boundary condition ${\bf g_i}$ and ${\bf g_i}$ separated by time $t$. The full dynamics of the system can be obtained, as in the particle case, by evaluating suitable quantum operators on the ensemble of geometries that contribute to the path integral. CDT describes how to compute this path integral, and how to choose the class of virtual paths. It also provides technical tools for extracting physical information about the quantum geometry.

As for the Regge calculus aspect, CDT uses the method in a manner different from its more classical applications. Rather than employ Regge calculus to approximate a spacetime, the aim of CDT is to approximate the path integral (3.0.2) as closely as possible, or rather to define it.

Note that CDT does not assume that the universe is constructed from 4-simplices; this Regge structure is only introduced in order to make the path integral manageable. The edge lengths are all fixed to a common value $a$, and the limit $a \to 0$ is eventually applied to the path integral to obtain its continuum limit, which is taken to be the actual physical prediction for the path integral of spacetime by the model. In order to achieve this continuum limit, the path integral is regularized: appropriate cut-off parameters for the configurations contributing to the path integral are introduced to make the path integral finite.

The key question becomes: how do we choose \emph{which} regularized triangulated geometries to select in computing the path integral?

Previous approaches using Euclidean "spacetimes" (where, of course, all dimensions are treated as spatial, with no distinction being given to a "time" direction) have failed; they result in a Hausdorff dimension of either $2$ or $\infty$ (for a 4-dimensional Euclidean "spacetime"), when the expected result would be 4 (however, according to some recent work (\cite{laiho}), Euclidean DT may be able to solve this problem by means of an additional parameter).

(A note on Hausdorff dimension: this is a measure obtained by comparing the typical linear size $r$ of a convex subspace of a given space (e.g. its diameter) with its volume $V(r)$. If the leading behaviour is $V(r) \sim r^{d_H}$, the space is said to have Hausdorff dimension $d_H$).

The requirement, then, is to find a path integral which allows for large short-scale fluctuations in curvature (accounting for the behaviour we expect to see at Planck scales, outlined at the start of this section), but in such a way that the resulting large-scale geometry does not degenerate completely (leading to a good classical limit). The apparent success of CDT in supplying this path integral lies in the imposition of causal rules on the building blocks; in effect, taking seriously the fact that a real spacetime is Lorentzian, not Euclidean, and as such contains some causal elements.

The nature of these causal rules is as such: each spacetime appearing in the sum over geometries ``\emph{should be a geometric object which can be obtained by evolving a purely spatial geometry in time, in such a way that its spatial topology is unchanged as a function of time}". Considerations of allowing some topology changes in time (obeying certain restrictions) have been made by the CDT authors, and an attempt to consider them will be made in the final section of this report.

\newpage

\section{The 2-D CDT Path Integral}

Now that the general strategy and goals of CDT have been presented, along with necessary tools obtained from Regge calculus, we can examine an application of the model. In general, obtaining results in CDT involves computer simulations and a large number of computations. The two-dimensional case, however, can be solved exactly, and offers some insight into the form taken by CDT results. The focus of the rest of this report is on the 2-D CDT path integral, following \cite{ambjorn98} very closely. First, the general structure of the 2-D triangulated spacetime is examined, and from there the full discrete path integral is constructed. Afterwards the coupling constants of the model are renormalized, and the continuum limit of the path integral is taken. Some interpretation of physical results is considered, after which some arguments involving topology changes are presented.

\subsection{Discrete Case}

The 2-D discrete CDT spacetime will have the form of closed 1-D spatial loops connected by triangles (Figure 3). The only geometric aspect of each spatial slice is its length, which is quantized in units of lattice spacing $a$, so that $L=l a$, where $l$ is an integer. Thus we will define spatial slices (in the discrete path integral) by $l$ vertices (or, equivalently, $l$ links connecting them). The 2-D geometry is formed by evolving an initial spatial loop in discrete time-steps, forming triangles.

The rules of propagation, as expressed in the source, are as follows: each vertex $i$ at time t is connected to $k_i$ vertices at time $t+1$, $k_i \geq 1$, by links with assigned length $ia$. To understand how the 2-D spacetime is formed, it helps to consider the process of evolving a spacetime in a step-by-step format:

\begin{itemize}
\item Choose number of points on entrance loop: $k$.
\item Choose number of points on exit loop: $l$.
\item Each point $i$ on the entrance is connected to $k_i \geq 1$ exit points.
\item Each set of these $k_i$ points forms $(k_i -1)$ lines on the exit loop \\ $\Rightarrow (k_i -1)$ triangles with their tip on the entrance loop per entrance point $i$.
\item Total number of triangles:
\begin{itemize}
\item $\sum_{i=1}^{k} (k_i -1) = l$ formed with their tip on the entrance loop (there are always $l$ triangles formed with their tip on the entrance loop)
\item $k$ triangles formed with their base on the entrance loop (formed between the $k$ points on the entrance loop)
\end{itemize}
$\Rightarrow$ Total triangles $= \sum_{i=1}^{k} (k_i -1) + k = \sum_{i=1}^{k} k_i -k +k = \sum_{i=1}^{k} k_i = k+l$
\item Finally, determine the number of ways to distribute the triangles for a given $k$ and $l$ (that is, the number of possible configurations that lead to different geometries). The way to calculate this number is presented a little later.
\end{itemize}

\begin{figure}
\begin{center}
\includegraphics[scale=0.5]{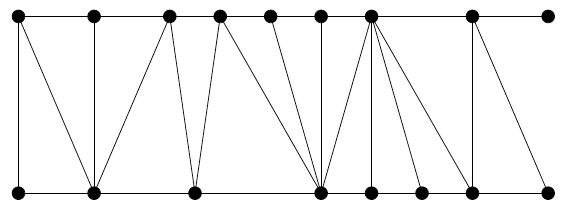} 
\end{center}
\caption{The propagation of a spatial slice from step $t$ to step $t+1$. The ends of the strip should be joined to form a band with topology $S 1 \times [0, 1]$. \\
\cite{ambjorn98}}
\end{figure}

So, the basic building blocks on this model are flat triangles with one space-like edge (length $a$) and two time-like edges (length $ia$). To understand where to go next, recall the final equation of the Regge calculus section:

\begin{equation*}
S_{Regge_{2 D}} = \kappa \, 2 \pi \chi - N(T) \lambda A.
\tag{2.0.17}
\end{equation*}

If we fix the topology ($\chi=2-2 \mathpzc{g}$ is constant), the first term contributes only a constant phase to the path integral; we therefore ignore it and focus only on the second term. Computing the area of the triangles (from the general formula for the area of an equilateral triangle):

\[
A=\frac{1}{2} a^2 \sqrt{\frac{\left( i a \right)^2}{a^2} - \frac{1}{4}} = \frac{1}{2} a^2 \sqrt{-1-\frac{1}{4}} = \frac{1}{2} a^2 i \sqrt{\frac{5}{4}} = i a^2 \frac{\sqrt{5}}{4}.
\]

Plugging into (2.0.17):

\[
S = - N(T) \lambda  i a^2 \frac{\sqrt{5}}{4} = \lambda a^2 N(T),
\]
where we have absorbed some constants into $\lambda$. Thus:

\begin{equation}
e^{i S_{CDT}} = e^{i \lambda a^2 N(T)}.
\end{equation}

The object of this section is to compute the path integral amplitude for propagating from a geometry of length $l_1$ to one of length $l_2$. This family of expressions will have an iterative form, and can be fully described by three equations:

\begin{equation}
G_{\lambda}^{(1)} (l_1,l_2) = \sum_{t=1}^{\infty} G_{\lambda}^{(1)}(l_1,l_2;t),
\end{equation}

\begin{equation}
G_{\lambda}^{(1)} (l_1,l_2;t) = \sum_{l=1}^{\infty} G_{\lambda}^{(1)}(l_1,l;1) \, l \, G_{\lambda}^{(1)}(l,l_2;t-1),
\end{equation}

\begin{equation}
G_{\lambda}^{(1)} (l_1,l_2;1) = \frac{1}{l_1} \sum_{\left\lbrace k_1,...,k_{l_1} \right\rbrace} e^{i \lambda a^2 \sum_{i=1}^{l_1} k_i}.
\end{equation}

The first equation is the amplitude for an arbitrary time separation of slices, derived by summing the second equation over $t$. The second equation reveals the iterative form the amplitude for a given $t$-value takes, and the third and final equation gives the base $t=1$ case. Note that the third expression has the form anticipated in (4.1.1), recalling that $N(T)$ represented the number of triangles in the spacetime, and $\sum_{i=1}^{k} k_i = k+l$ indeed was shown earlier to represent the total number of triangles.

It is important here to make a note about vertex marking. The above expressions give the physical amplitude for the cases explained in the previous paragraph. Marking a vertex on a loop corresponds to multiplication by the loop length factor $l$. This is because the marking of a loop means that cyclic permutations of identical triangulations are included in the summation (one may think of this as looking at the same configuration from each of $l$ vertices on the marked loop). In (4.1.4), the curly-bracket sum includes cyclic permutations of the entrance loop (which has $l_1$ vertices), and thus a factor of $l_1$ appears for each unique way of triangulating the spacetime (it will be shown later, when the combinatoric factor $\mathcal{N}(k,l)$ is calculated, that the curly-bracket sum indeed includes cyclic permutations). These permutations, however, do not alter the geometry of the configuration, and thus do not make any physical contribution. It is for this reason that the $1/ l_1$ factor appears in the third equation: it is there to eliminate the cyclic permutations and give the physical amplitude.

Throughout the rest of the 2-D CDT path integral derivation in the source (in both the discrete and continuous parts), vertices on loops will be occasionally marked and unmarked in order to simplify calculations with the understanding that the end results can usually be modified to give any state of marking by multiplication or division by $l$s (or $L$s in the continuum limit). The form predominantly used (and which will be adopted for this report) is:

\begin{equation}
G_{\Lambda}(l_1,l_2;t) \equiv l_1 G^{(1)}_{\Lambda}(l_1,l_2;t),
\end{equation}
where a vertex on the entrance loop has been marked to get rid of the factors of $l$ and $1/l_1$ in (4.1.3) and (4.1.4).

$G_\lambda(l_1,l_2;t)$ plays the role of a transfer matrix, obeying the properties:

\begin{equation}
G_\lambda(l_1,l_2;t_1+t_2) = \sum_{l} G_\lambda(l_1,l;t_1) G_\lambda(l,l_2;t_2),
\end{equation}

\begin{equation}
G_\lambda(l_1,l_2;t+1) = \sum_{l} G_\lambda(l_1,l;1) G_\lambda(l,l_2;t).
\end{equation}

For convenience, the generating function for the $G_\lambda(l_1,l_2;t)$'s is introduced:

\begin{equation}
G_\lambda(x,y;t) \equiv \sum_{k,l} x^k y^l G_\lambda(k,l;t).
\end{equation}

We can use the above formula to rewrite (4.1.6) in terms of $x$ and $y$:

\begin{equation}
G_{\lambda}(x,y;t_{1}+t_{2}) = \oint{\frac{dz}{2 \pi i z} G_{\lambda}(x,z^{-1};t_{1}) G_{\lambda}(z,y;t_{2})},
\end{equation}
where the contour is chosen not to include the singularities of $G_{\lambda}(z,y;t_{2})$.

\begin{center}
\fbox{
\begin{minipage}[b]{0.8\linewidth}
Proof that:

\begin{equation*}
G_{\lambda}(x,y;t_{1}+t_{2}) = \oint{\frac{dz}{2 \pi i z} G_{\lambda}(x,z^{-1};t_{1}) G_{\lambda}(z,y;t_{2})}
\tag{4.1.9}
\end{equation*}.

corresponds to the combination of the property:
\begin{equation*}
G_{\lambda}(k,l;t_{1}+t_{2}) = \sum_{m} G_{\lambda}(k,m;t_{1}) G_{\lambda}(m,l;t_{2})
\tag{4.1.6}
\end{equation*}

with the generating function:
\begin{equation*}
G_{\lambda}(x,y;t) \equiv \sum_{k,l} x^{k} y^{l} G_{\lambda}(k,l;t).
\tag{4.1.8}
\end{equation*}

Replacing the two $G_{\lambda}$'s in (4.1.9) with the form given by (4.1.8) we have:

\[		
G_{\lambda}(x,y;t_{1}+t_{2}) = \oint{\frac{dz}{2 \pi i z} \sum_{k,m} x^{k} \left(	\frac{1}{z} \right)^{m} G_{\lambda}(k,m;t_{1}) \sum_{n,l} z^{n} y^{l} G_{\lambda}(n,l;t_{2})}
\]

\[
= \sum_{k,m} \sum_{n,l} x^{k} y^{l} \left( \oint{\frac{dz}{2 \pi i z} z^{n-m}} \right) G_{\lambda}(k,m;t_{1}) G_{\lambda}(n,l;t_{2}).
\]

Using a change of coordinates to the effect $z=r e^{i \phi}$:

\[
\oint{\frac{dz}{2 \pi i z} z^{n-m}} = \frac{1}{2 \pi i} \int_0^{2\pi} {\frac{i r e^{i \phi} d\phi}{r e^{i \phi}} \left( r e^{i \phi} \right)^{n-m}}
\]

\[
= \frac{1}{2 \pi} r^{n-m} \int_0^{2\pi}{d\phi e^{i \phi (n-m)}}
\]

\[
= \frac{1}{2 \pi} 2 \pi \delta_{n m}
\]

\end{minipage}
}

\fbox{
\begin{minipage}[b]{0.8\linewidth}

\[
\Rightarrow G_{\lambda}(x,y;t_{1}+t_{2}) = \sum_{k,m} \sum_{n,l} x^{k} y^{l} \delta_{n m} G_{\lambda}(k,m;t_{1}) G_{\lambda}(n,l;t_{2})
\]

\[ 
= \sum_{k,l,m} x^{k} y^{l} G_{\lambda}(k,m;t_{1}) G_{\lambda}(m,l;t_{2})
\]

\[
= \sum_{k,l} x^{k} y^{l} \left( \sum_{m} G_{\lambda}(k,m;t_{1}) G_{\lambda}(m,l;t_{2}) \right).
\]

Recognizing the term in brackets as the R.H.S. of (4.1.6), we conclude:

\[
G_{\lambda}(x,y;t_{1}+t_{2}) = \sum_{k,l} x^{k} y^{l} G_{\lambda}(k,l;t_{1}+t_{2}).
\]

This is precisely the form of (4.1.8) for $t=t_{1}+t_{2}$. Q.E.D.

\end{minipage}
}
\end{center}

To obtain the path integral:

\[
\begin{split}
G_\lambda(x,y;1) &= \sum_{k,l} x^k y^l G_\lambda(k,l;1) \\
&= \sum_{k,l} x^k y^l \sum_{\left\lbrace k_1,...,k_k \right\rbrace} e^{i \lambda a^2 \sum_{i=1}^k k_1} \\
&= \sum_{k,l} x^k y^l \sum_{\left\lbrace k_1,...,k_k \right\rbrace} g^{k+l} \quad \left( g \equiv e^{i \lambda a^2}, \; \sum_{i=1}^{k} k_i = k+l \right)
\end{split}
\]

\begin{equation}
\Rightarrow G(x,y;g;1) \equiv G_\lambda(x,y;1) = \sum_{k,l} (g x)^k (g y)^l \mathcal{N}(k,l).
\end{equation}

The summation over $\left\lbrace k_1 ,...,k_2 \right\rbrace$ indicates a summation over all possible distributions of values of $k_1 ,...,k_2$. Since each of these distributions will contribute a factor of $g^{k+l}$ (the total number of triangles is restricted to $k+l$ at all times), we can express the summation as a multiplication by $\mathcal{N}(k,l)$, which is a combinatoric factor counting all possible distributions as described.

Based on what was discussed earlier, we expect $\mathcal{N}(k,l)$ to give the number of ways of distributing $k+l$ triangles between the points of two consecutive spatial slices (Figure 3 illustrates this well) such that no point is left unconnected (recall $k_i \geq 1$).

It turns out that this imposition (that no point is left unconnected) means that we can assign $k$ triangles with their tip on the exit loop and just find out how to assign the remaining $l$ triangles (which have their tip on the entrance loop). Allowing for maximal freedom in distributing the $l$ triangles corresponds to there only being one way of distributing the $k$ triangles (changing the placement of one of the $k$ triangles always corresponds to changing the placement of at least one of the $l$ triangles, and this possibility is accounted for in the various configurations of the $l$ triangles).

\begin{figure}
\begin{center}
\includegraphics[scale=0.5]{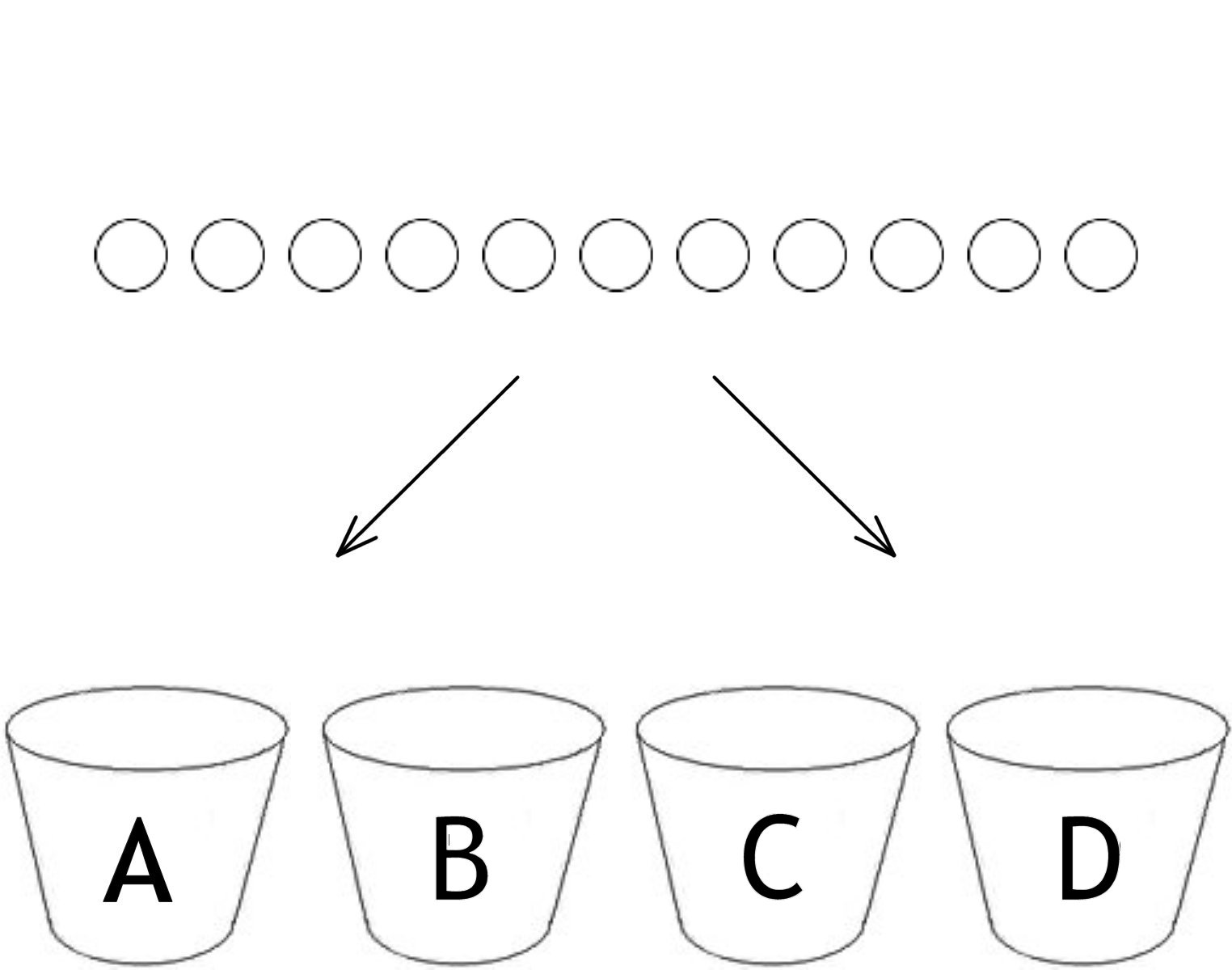} 
\end{center}
\caption{Distributing indistinguishable balls into labelled buckets.}
\end{figure}

The problem then is finding how many ways there are to distribute $l$ triangles with their tips somewhere in the set of $k$ entrance loop points. This translates into the standard combinatorial problem of distributing $l$ indistinguishable balls into $k$ labelled buckets. The situation is illustrated in Figure 4. A simpler way is to think of the situation as a problem of how many ways there are to order a string of $(l+k-1)$ objects, where $l$ objects are of one type and $(k-1)$ objects are of another type. The $(k-1)$ objects are the separators formed by the $k$ buckets, and they are understood to be indistinguishable (it does not matter \emph{which} border is first or second or third, but instead how many of the $l$ balls are to the left and right of it).

Total ways to permute the string:

\begin{equation}
\mathcal{N}(k,l) = \frac{(l+k-1)!}{l!(k-1)!},
\end{equation}
representing the fact that there are $(l+k-1)$ items being shuffled amongst themselves, but $l$ of them are identical to each other and the other $(k-1)$ are also identical to each other (so we must divide by the number of ways to order each of them to arrive at the true number of unique configurations).

Note: The $\mathcal{N}(k,l)$ formula includes cyclic permutations of the entrance loop points (or, in the terms of our analogy, the buckets). Since these do not produce distinct geometries, our final expression for $G^{(1)}_{\Lambda}(k,l;1)$ needs a factor of $1/k$ to correct that, as is reflected in (4.1.4). Note that $\frac{1}{k} \mathcal{N}(k,l)=\frac{(l+k-1)!}{l!k!}$, which is symmetric in $(k,l)$. This is the overall factor that appears in $G^{(1)}_{\Lambda}(k,l;1)$, which (as argued earlier) corresponds to the physical propagator for our discrete model. The propagator we work with is actually (4.1.5), which is not symmetric in $(k,l)$ and is not the physical path integral (but one may always retrieve the physical one by dividing by $k$).

Thus, returning to (4.1.10) and making the sums explicit, we have:

\[
\begin{split}
G(x,y;g;1) = \sum_{k=0}^{\infty} \left[ {\left( g x \right)}^k \underbrace{\sum_{l=0}^{\infty} {\left( g y \right)}^l \frac{(l+k-1)!}{l! (k-1)!}} \right]. \\
= \frac{1}{(1-g y)^k}
\end{split}
\]

\begin{center}
\fbox{
\begin{minipage}[b]{0.8\linewidth}
Proof that:

\begin{equation}
\sum_{l=0}^{\infty} {\left( z \right)}^l \frac{(l+k-1)!}{l! (k-1)!} = \frac{1}{(1-z)^k}.
\end{equation}

Beginning with the R.H.S., observe that $\frac{1}{(1-z)^k}$ can be written as the $(k-1)$th derivative of $\frac{1}{1-z}$, multiplied by some factor:

\[
\frac{d}{dz} \frac{1}{1-z} = \frac{1}{(1-z)^2}, \quad \frac{d}{dz} \frac{1}{(1-z)^2} = \frac{2}{(1-z)^3}
\]

\[
\Rightarrow \frac{d^{k-1}}{d z^{k-1}} \left( \frac{1}{1-z} \right) = \frac{(k-1)!}{(1-z)^k}
\]

\begin{equation*}
\Rightarrow \frac{1}{(1-z)^k} = \frac{1}{(k-1)!} \frac{d^{k-1}}{d z^{k-1}} \left( \frac{1}{1-z} \right).
\tag{A}
\end{equation*}

Now, expanding the geometric series:

\[
\frac{d^{k-1}}{d z^{k-1}} \left( \frac{1}{1-z} \right) = \frac{d^{k-1}}{d z^{k-1}} \left[ \sum_{i=0}^{\infty} z^i \right].
\]

For some $i=n \geq k-1$ (for $n < k-1$ the term vanishes):

\begin{equation*}
\frac{d^{k-1}}{d z^{k-1}} z^n = C z^{n-k+1},
\tag{B}
\end{equation*}

where C has the form:

\[
\begin{split}
C &= n(n-1)(n-2)...(n-k+2) \\
&= \frac{n(n-1)(n-2)...(n-k+2)(n-k+1)(n-k)...2 \cdot 1}{(n-k+1)(n-k)...2 \cdot 1} \\
&= \frac{n!}{(n-k+1)!}.
\end{split}
\]

\end{minipage}
}

\end{center}

\begin{center}
\fbox{
\begin{minipage}[b]{0.8\linewidth}

Define $l \equiv n-k+1$ (this is our new counter with limits $0 \to \infty$ in the final, differentiated result).

\[
\Rightarrow C = \frac{(l+k-1)!}{l!}.
\]

Inserting this result into (B) and (A), we have:

\begin{equation*}
\frac{1}{(1-z)^k} = \sum_{l=0}^{\infty} z^l  \frac{(l+k-1)!}{l! (k-1)!}
\tag{4.1.12}.
\end{equation*}

\end{minipage}
}

\end{center}

Continuing with our original derivation, we have:

\[
\begin{split}
G(x,y;g;1) &= \sum_{k=0}^{\infty} \left[ {\left( g x \right)}^k \frac{1}{(1-g y)^k} \right] \\
&= \sum_{k=0}^{\infty} \left[ {\left( g x \right)}^k {\left( \sum_{l=0}^{\infty} {\left( g y \right)}^l \right)}^k \right] \\
&= \sum_{k=0}^{\infty} { \left( g x \sum_{l=0}^{\infty} {\left( g y \right)}^l \right)}^k - \underbrace{\sum_{k=0}^{\infty} {\left( g x \right)}^k}.
\end{split}
\]

The term with the underbrace has been added in order to exclude the contribution of $l=0$, since that corresponds to a singularity. In this manner the also singular $k=0$ case also makes no contribution to our path integral.

To get the final expression without summations:

\[
\begin{split}
G(x,y;g;1) &= \sum_{k=0}^{\infty} { \left( g x \frac{1}{1-g y} \right)}^k - \sum_{k=0}^{\infty} {\left( g x \right)}^k \\
&= \frac{1}{1-\frac{g x}{1-g y}} - \frac{1}{1-g x} \\
&= \frac{1-g y}{1-g x-g y} - \frac{1}{1-g x} \\
&= \frac{(1-g y)(1-g x)-1+g x+g y}{(1-g x)(1-g x-g y)}
\end{split}
\]

\begin{equation}
\Rightarrow G(x,y;g;1) = \frac{g^2 x y}{(1-g x)(1-g x-g y)}.
\end{equation}

Combining (4.1.13) with (4.1.9), with $t_1=1$:

\[
\begin{split}
G_{\lambda}(x,y;1+\left( t-1 \right)) &= \oint{\frac{dz}{2 \pi i z} G_{\lambda}(x,z^{-1};1) G_{\lambda}(z,y;t-1)} \\
&= \oint{\frac{dz}{2 \pi i} \frac{\frac{g^2 x}{z}}{z \left( 1-g x \right) \left( 1-g x-\frac{g}{z} \right)} G_{\lambda}(z,y;t-1)} \\
&= \oint{\frac{dz}{2 \pi i} \frac{g^2 x}{z (1-g x) (z (1-g x)-g)} G_{\lambda}(z,y;t-1)}.
\end{split}
\]

The integrand has poles at $z_0=0$ and $z_1=\frac{g}{1-g x}$. \\*

To compute the integral, we use the residue theorem:

\begin{equation}
\oint_{\gamma} f(z) dz = 2 \pi i \sum_{z_k \in \gamma} {\rm Res}(f;z_k)
\end{equation}
along with the property:

\begin{equation}
{\rm Res} \left( \frac{F}{G}; z_1 \right) = \frac{F(z_1)}{G^\prime (z_1)}.
\end{equation}

In our case:

\[
F(z_1)=\frac{g^2 x G_{\lambda}(z,y;t-1)}{1-g x}, \quad G(z_1)=z (z(1-g x)-g) = z^2 (1-g x) - g z
\]
(the $z_0$ pole does not contribute because $G_{\lambda}(z_0=0,y;t-1)=0$).

\[
\Rightarrow G^\prime = 2 z (1-g x) - g = 2 g - g = g
\]

\[
\begin{split}
\Rightarrow Res \left( \frac{F}{G}; z_1 \right) &= \frac{g^2 x}{1-g x} \frac{1}{g} G_{\lambda} \left( \frac{g}{1-g x},y;t-1 \right) \\
&= \frac{g x}{1-g x} G_{\lambda} \left( \frac{g}{1-g x},y;t-1 \right)
\end{split}
\]

\begin{equation}
\Rightarrow G(x,y;g;t) = \frac{g x}{1-g x} G \left( \frac{g}{1-g x},y;g;t-1 \right).
\end{equation}

We can iterate this expression and write the solution:

\begin{equation}
G(x,y;g;t) = F_1^2(x) F_2^2(x)...F_{t-1}^2(x) \frac{g^2 x y}{\left[ 1-g F_{t-1}(x) \right] \left[ 1-g F_{t-1}(x)-g y \right]},
\end{equation}
where $F_t(x)$ is defined iteratively by:

\begin{equation}
F_t(x) = \frac{g}{1-g F_{t-1}(x)}, \quad F_0(x)=x.
\end{equation}
We can verify this by examining $G(x,y;g;2)$:

\[
\begin{split}
G(x,y;g;2) &= \frac{g x}{1-g x} G \left( \frac{g}{1-g x},y;g;1 \right) \\
&= \frac{g x}{1-g x} \left( \frac{g}{1-g x} \right) \frac{g^2 y}{\left( 1-\frac{g^2}{1-g x} \right) \left( 1- \frac{g^2}{1-g x} -g y \right)}
\\
&= \left( \frac{g}{1-g x} \right)^2 \frac{g^2 x y}{\left( 1- \frac{g^2}{1-g x} \right) \left( 1- \frac{g^2}{1-g x} - g y \right) } \\
&= F_1^2(x) \frac{g^2 x y}{\left( 1- g F_1(x) \right) \left( 1- g F_1(x) - g y \right)}.
\end{split}
\]
Continuing in this manner,

\[
\begin{split}
G(x,y;g;3) &= F_1^2(F_1(x)) \frac{g x}{1-g x} \frac{g}{1-g x} \frac{g^2 y}{\left[ 1- g F_1(F_1(x)) \right] \left[ 1- g F_1(F_1(x)) - g y \right]} \\
&= F_1^2(x) F_2^2(x) \frac{g^2 x y}{\left[ 1- g F_2(x) \right] \left[ 1- g F_2(x) - g y \right]},
\end{split}
\]
and we can thus see that (4.1.17) holds.

If we express (4.1.17) in terms of a fixed point, we can write it in the form:

\begin{equation}
F_t(x) = F \frac{1-x F +F^{2 t -1}(x-F)}{1-x F +F^{2 t +1}(x-F)}, \quad F=\frac{1-\sqrt{1-4 g^2}}{2 g}.
\end{equation}
The fixed point F is found by solving $F=\frac{g}{1-g F}$:

\[
\Rightarrow F - g F^2 - g = 0 \rightarrow F = \frac{1 \pm \sqrt{1-4 g^2}}{2 g}.
\]
Note: The authors choose ``$-$" sign in the fixed point formula. The justification for this appears to be that choosing either solution leads to no loss of generality, since both correspond to the same value of the physical constant $g$:

\begin{equation}
g=\frac{F}{1+F^2}.
\end{equation}

We may verify that (4.1.17) and (4.1.19) are in correspondence by considering $F_{t-1}(x)$ as defined by (4.1.19):

\[
F_{t-1}(x) = F \frac{1-x F +F^{2 t -3}(x-F)}{1-x F +F^{2 t -1}(x-F)}
\]

\[
\Rightarrow \frac{g}{1-g F_{t-1}(x)} = \frac{g(1-x F +F^{2 t-1}(x-F))}{(1-x F +F^{2 t-1}(x-F))-g F (1-x F +F^{2 t-3}(x-F))}.
\]

Expressing $g$ in terms of $F$ as per (4.1.20) and performing some straightforward algebra, we obtain:

\[
\frac{g}{1-g F_{t-1}(x)} = F \left( \frac{1-x F +F^{2 t-1}(x-F)}{1-x F +F^{2 t+1}(x-F)} \right).
\]
We recognize the above expression as $F_t(x)$. Therefore:

\[
F_t(x) = \frac{g}{1-g F_{t-1}(x)},
\]
which is our iterative expression (4.1.18) (the base case $F_0(x)=x$ can be readily verified to correspond to (4.1.19) by setting $t=0$ and simplifying).

We can use this expression to express $G(x,y;g;t)$ (as defined by (4.1.17)) in terms of the fixed point and obtain:

\begin{equation}
G(x,y;g;t) = \frac{F^{2 t} (1-F^2)^2 x y}{(A_t-B_t x) (A_t-B_t(x+y)+C_t x y)},
\end{equation}
where the time-dependent coefficients are given by:

\begin{equation}
A_t = 1-F^{2 t+2}, \quad B_t=F (1-F^{2 t}), \quad C_t=F^2 (1-F^{2 t-2}).
\end{equation}
Note that, by substituting in the time-dependent coefficients and performing some simplifications, (4.1.21) can be written as:

\begin{equation}
G(x,y;g;t) = \frac{F^{2 t} (1-F^2)^2 x y}{[1-x F-F^{2 t+1} (F-x)] [(1-x F)(1-y F)-F^{2 t}(F-x)(F-y)]}.
\end{equation}

Expanding $F_1(x) F_2(x) F_3(x) \ldots F_{t-1}(x)$ using (4.1.19):

\[
F \frac{1-x F+F (x-F)}{\cancel{1-x F+F^3 (x-F)}} \cdot F \frac{\cancel{1-x F+F^3 (x-F)}}{\cancel{1-x F+F^5 (x-F)}} \cdot F \frac{\cancel{1-x F+F^5 (x-F)}}{1-x F+F^7 (x-F)} \ldots
\]
We can clearly see by inspection that the only surviving terms of this series will be:

\[
F^{t-1} \frac{1-x F+F (x-F)}{1-x F+F^{2 (t-1)+1} (x-F)}.
\]
Thus:

\begin{equation}
F_1(x) F_2(x) \ldots F_{t-1}(x) = F^{t-1} \frac{1-F^2}{1-x F+F^{2 t-1} (x-F)}.
\end{equation}

Now:

\[ 1-g F_{t-1}(x)=1-g F \frac{1-x F+F^{2 t-3} (x-F)}{1-x F+F^{2 t-1} (x-F)}
\]

\[
= 1-\frac{F^2}{1+F^2} \frac{1-x F+F^{2 t-3} (x-F)}{1-x F+F^{2 t-1} (x-F)} \quad \left( g=\frac{F}{1+F^2} \right).
\]
After some algebra we find that:

\begin{equation}
1-g F_{t-1}(x)=\frac{1-x F-F^{2 t+1} (F-x)}{(1+F^2)(1-x F +F^{2 t-1} (x-F))} \equiv \frac{M}{A}.
\end{equation}
Similar manipulations give us:

\begin{equation}
1-g F_{t-1}(x)-g y=\frac{(1-x F)(1-y F)-F^{2 t}(F-x)(F-y)}{(1+F^2)(1-x F +F^{2 t-1} (x-F))} \equiv \frac{N}{A}.
\end{equation}

Substituting (4.1.24), (4.1.25) and (4.1.26) into (4.1.17):

\[
G(x,y;g;t) = F^{2 t-2} \frac{(1-F^2)^2}{(1-x F+F^{2 t-1} (x-F))^2} \frac{F^2}{(1+F^2)^2} \frac{x y}{M N} A^2
\]

\[
= F^{2 t-2} \frac{(1-F^2)^2}{\cancel{(1-x F+F^{2 t-1} (x-F))^2}} \frac{F^2}{\cancel{(1+F^2)^2}} \frac{x y}{M N} \cancel{(1+F^2)^2}\cancel{(1-x F +F^{2 t-1} (x-F))^2}
\]

\[
=\frac{F^{2 t} (1-F^2)^2 x y}{M N}.
\]

Going back to (4.1.23), one recognizes M and N as being the two terms in square brackets in the denominator. Therefore:

\[
G(x,y;g;t) = \frac{F^{2 t} (1-F^2)^2 x y}{[1-x F-F^{2 t+1} (F-x)] [(1-x F)(1-y F)-F^{2 t}(F-x)(F-y)]}
\]
as claimed.

The final goal is to compute $G_\lambda(l_1,l_2;t)$ from $G(x,y;g;t)$ (recall that $l_1$, $l_2$ are measures of the length of the loops; $x$ and $y$ have no direct physical meaning). This can be achieved by means of a discrete inverse Laplace transformation:

\begin{equation}
G_{\lambda}(k,l;t) =  \oint{\frac{dx}{2 \pi i x}} \oint{\frac{dy}{2 \pi i y}} \frac{1}{x^{k}} \frac{1}{y^{l}} G_{\lambda}(x,y;t).
\end{equation}

This relation can be derived from the generating function, defined by (4.1.8):

\[
G_{\lambda}(x,y;t) = \sum_{k,l} x^{k} y^{l} G_{\lambda}(k,l;t).
\]

Dividing both sides by $2 \pi i x x^p$ and $2 \pi i y y^q$:

\[
\Rightarrow \frac{1}{2 \pi i x} \frac{1}{2 \pi i y} \frac{1}{x^p} \frac{1}{y^q} G_{\lambda}(x,y;t) = \sum_{k,l} \frac{1}{2 \pi i x} \frac{1}{2 \pi i y} x^{k-p} y^{l-q} G_{\lambda}(k,l;t).
\]

Performing a loop integral enclosing no singularities on both sides:
\begin{equation}
\Rightarrow \oint{\frac{dx}{2 \pi i x}} \oint{\frac{1}{2 \pi i y}} \frac{1}{x^p} \frac{1}{y^q} G_{\lambda}(x,y;t) = \sum_{k,l} \left( \oint{\frac{dx}{2 \pi i x} x^{k-p}} \right) \left( \oint{\frac{dy}{2 \pi i y} y^{l-q}} \right) G_{\lambda}(k,l;t).
\end{equation}

Applying a coordinate change $z=r e^{i \phi}$ as before to the first term in brackets:

\[
\oint{\frac{dx}{2 \pi i x} x^{k-p}} = \frac{i}{2 \pi i} r^{1+k-p-1} \int_0^{2 \pi} d\phi e^{i \phi (k-p)}
\]

\[
= \frac{1}{2 \pi} r^{k-p} 2 \pi \delta_{k p}
\]

\begin{equation}
\Rightarrow \oint{\frac{dx}{2 \pi i x} x^{k-p}} = \delta_{k p}.
\end{equation}

The second term in brackets gives $\delta_{l q}$ by the same method and, returning to (4.1.28) we have:

\[
\oint{\frac{dx}{2 \pi i x}} \oint{\frac{1}{2 \pi i y}} \frac{1}{x^p} \frac{1}{y^q} G_{\lambda}(x,y;t) = \sum_{k,l} \delta_{k p} \delta_{l q} G_{\lambda}(k,l;t)
\]

\[
\Rightarrow G_{\lambda}(p,q;t) =  \oint{\frac{dx}{2 \pi i x}} \oint{\frac{dy}{2 \pi i y}} \frac{1}{x^{p}} \frac{1}{y^{q}} G_{\lambda}(x,y;t).
\]

An alternative method, used by the authors, is to rewrite (4.1.21) as a power series in $x$ and $y$, with the final result for the discrete 2D CDT path integral being:

\begin{equation}
G_{\lambda}(l_1,l_2;t) = \frac{F^{2 t} (1-F^2)^2 B_t^{l_1+l_2}}{l_2 A_t^{l_1+l_2+2}} \sum_{k=0}^{{\rm min}(l_1,l_2)-1} \frac{l_1+l_2-k-1}{k! (l_1-k-1)! (l_2-k-1)!} \left( \frac{A_t C_t}{B_t^2} \right)^k.
\end{equation}

\newpage

\subsection{The Continuum Limit}

Based on the similarity between our path integral formalism used in the previous section and the usual representation of a particle as a sum over free paths, we expect to be able to apply the renormalization  methods of the latter in order to obtain the continuum limit of our discrete path integral.

As such, we expect an additive normalization for our positive mass dimension coupling constants (i.e. the cosmological constant and the ``boundary cosmological constants"):

\begin{equation}
\lambda = \frac{C_\lambda}{a^2} + \tilde{\Lambda}, \quad \lambda_i = \frac{C_{\lambda_{i}}}{a} + \tilde{X}, \quad \lambda_o = \frac{C_{\lambda_{o}}}{a} + \tilde{Y},
\end{equation}
where $\tilde{\Lambda}, \tilde{X}, \tilde{Y}$ are the renormalized cosmological and boundary cosmological constants. Defining:

\begin{equation}
g_c = e^{i C_\lambda}, \quad x_c = e^{i C_{\lambda_i}}, \quad y_c = e^{i C_{\lambda_o}},
\end{equation}
it follows from our definitions of g, x and y in the previous section that:

\begin{equation}
g = g_c e^{i a^2 \tilde{\Lambda}}, \quad x = x_c e^{i a \tilde{X}}, \quad y = y_c e^{i a \tilde{Y}}.
\end{equation}

The renormalization of the path integral, again analogously to the free particle case, appears as a multiplicative factor:

\begin{equation}
G_{\tilde{\Lambda}}(\tilde{X},\tilde{Y};T) = \lim_{a\to 0} a^{\eta} G(x,y;g;t).
\end{equation}

The claim is that the only possible choice of $\eta$ for which the right-hand side survives the limit is $\eta=1$. To show this, we return to a previously seen property of $G_{\lambda}$ and consider the form of its continuum limit:

\begin{equation}
G_{\lambda}(k,l;t_{1}+t_{2}) = \sum_{m} G_{\lambda}(k,m;t_{1}) G_{\lambda}(m,l;t_{2}).
\end{equation}

Expressing this in the shorthand form $\sum_{m}f(m)$ and performing some manipulations:

\[
\sum_{m}f(m) = \frac{1}{a} \left(\sum_{m}f(m) a\right) = \frac{1}{a} \left(\sum_{m}f(m) \Delta l\right) \to \frac{1}{a} \int dM f(M), \quad M=m a.
\]
Thus we have:

\begin{equation}
G_{\lambda}(k,l;t_{1}+t_{2}) \to \frac{1}{a} \int d M G_{\lambda}(k,M;t_{1}) G_{\lambda}(M,l;t_{2}).
\end{equation}

We require this to survive the $a \to 0$ limit, therefore we need both sides to be of same order in a. By inspection the only way to satisfy this is if:

\begin{equation}
G_{\lambda}(k,l;t) \propto a, \quad a \to 0.
\end{equation}
It will be shown later on that this is indeed the case.

Now let us consider our expression for the generating function for $G_{\lambda}(k,l;t)$:

\begin{equation}
G_{\lambda}(x,y;t) = \sum_{k,l} x^k y^l G_{\lambda}(k,l;t).
\end{equation}
Applying the process shown above but with two sums, we have:

\[
G_{\lambda}(x,y;t) \to \frac{1}{a^2} \iint d K d L x^{\frac{K}{a}} y^{\frac{L}{a}} G_{\lambda}(K,L;t).
\]
Thus we have:

\begin{equation}
G_{\lambda}(x,y;t) \propto \frac{1}{a^2} a = \frac{1}{a}, \quad a \to 0.
\end{equation}
Therefore, in expression (4.2.4), we would need $\eta=1$ so that the right-hand side is of order 0 in a as $a \to 0$.

In the paper it is argued that only for particular values of the dynamic variables will we obtain a non-trivial continuum limit. The authors choose $\lvert F\rvert \to 1$. The reason for this choice seems to be getting rid of the $t$-dependence of $G(x,y;g;t)$; however, the exact reasons for this were not made obvious. This leads to $F = e^{i \alpha}$ and thus, from our earlier relation between $g$ and $F$:

\begin{equation}
g_{c} = \frac{1}{2 \cos{\alpha}} \quad {\rm for} \quad F = e^{i \alpha}, \quad \alpha \in \Re \qquad ({\rm as} \: g \to g_{c}).
\end{equation}

The only interesting choices for $g_{c}$ are stated to be $g_{c} = \pm 1/2$, and the authors choose $g_{c} = 1/2$ without loss of generality (explained in the next step). The reasons for this choice are also unclear and warrant further study. It is also determined that, given this choice of $g_{c}$, the only interesting choices for $x$ and $y$ are $x,y \to 1$ (choosing $g_{c} = -1/2$ in the previous step leads to the choice $x,y \to -1$ and the end result is unchanged). Thus $x_{c},y_{c} \to 1$.

Note that the choices of values made by the authors are justified to an extent, since they produce a reasonable-looking continuum limit. What makes the reasons for the choices worth studying is the possibility of there being multiple ways of choosing limits, and the question of whether or not they result in the same continuum limit for $G(x,y;g;t)$. 

In order to approach the above values from the region where $G(x,y;g;t)$ converges, the renormalized coupling $\tilde{\Lambda}$ is chosen to also be imaginary $( \tilde{\Lambda}=i\Lambda )$. The same argument is followed for the so-called ``boundary cosmological constants": $\tilde{X}=i X, \tilde{Y}=i Y$. This leaves us with our original values $g$, $x$ and $y$ taking the form:

\[
g=\frac{1}{2} e^{-\Lambda a^2}, \quad x=e^{-X a}, \quad y=e^{-Y a}.
\]

An important note to make is that at this stage we are talking about the Euclidean sector of the theory; the transition to the Lorentzian form is obtained by employing the analytic continuation $\Lambda \to -i \Lambda$, where $\Lambda$ is our renormalized coupling constant. However, the authors point out that there is a crucial distinction between this model and older Euclidean triangulation models: our choice of only geometries admitting a causal structure at the stage of our triangulation construction rules would not have been well justified in a purely Euclidean model (since the notion of causality implies a dimension of time, distinct from the rest).

Summarizing the results of our choices of limits, we have:

\begin{equation}
g=\frac{1}{2} e^{-\Lambda a^2} \to \frac{1}{2} \left( 1 - \frac{1}{2} \Lambda a^2 \right), \quad ({\rm i.e.} \; F \to 1-a\sqrt{\Lambda} \approx e^{-a \sqrt{\Lambda}}),
\end{equation}

\begin{equation}
x=e^{-X a} \approx 1-a X, \qquad y=e^{-Y a} \approx 1-a Y,
\end{equation}
where the arrows in the first set of relations signify a redefinition of the coupling constant $\Lambda$, performed to get rid of factors of $1/2$ and such in the upcoming formulae. In the second set we are taking the first two terms of a Taylor expansion, given that we are taking the $a \to 0$ limit.

The object here is substitute these expressions into (4.1.23) (the form of $G(x,y;g;t)$ derived earlier) and simplify. Note that throughout the derivation the fact that we are taking $a \to 0$ will be used to ignore next-to-leading-order terms and to express exponentials in $a$ in the form given in (4.2.12). For clarity we will split $G(x,y;g;t)$ into parts as such:

\begin{equation}
G(x,y;g;t) = \frac{A}{B C},
\end{equation}

\begin{equation}
A \equiv F^{2 t} (1-F^2)^2 x y, \quad B \equiv A_t-B_t x, \quad C \equiv A_t-B_t(x+y)+C_t x y,
\end{equation}
where $A_t, B_t$ and $C_t$ were defined earlier. Plugging (4.2.11) into the expression for A, we have:

\[
A = e^{-2 t a \sqrt{\Lambda}} (1-e^{-2 a \sqrt{\Lambda}})^2 (1-a X) (1-a Y).
\]
Using $T=t a$, we have:

\[
A = e^{-2 \sqrt{\Lambda} T} (1-(1-2 a \sqrt{\Lambda}))^2 (1-a X) (1-a Y)
\]

\[
\Rightarrow A = e^{-2 \sqrt{\Lambda} T} 4 a^2 \Lambda (1-a X) (1-a Y).
\]
Dropping terms next-to-leading-order in a, we have:

\begin{equation}
A = 4 \Lambda a^2 e^{-2 \sqrt{\Lambda} T}.
\end{equation}
Now for B:

\[
B = 1 - {(e^{-a \sqrt{\Lambda}})}^{2 t +2} - e^{-a \sqrt{\Lambda}} (1-e^{-2 t a \sqrt{\Lambda}}) (1-a X)
\]

\[
= 1 - (e^{-2 \sqrt{\Lambda} T}) (e^{-2 a \sqrt{\Lambda}}) - e^{-a \sqrt{\Lambda}} (1-e^{-2 \sqrt{\Lambda} T}) (1-a X)
\]

\[
=e^{-2 \sqrt{\Lambda} T} \left( -e^{-2 a \sqrt{\Lambda}} +e^{-a \sqrt{\Lambda}} (1-a X) \right) + \left( 1 - e^{-a \sqrt{\Lambda}} (1-a X) \right).
\]
Writing $e^{-C a \sqrt{\Lambda}} \approx 1 - C a \sqrt{\Lambda}$:

\[
B = e^{-2 \sqrt{\Lambda} T} \left( -1 + 2 a \sqrt{\Lambda} + (1-a \sqrt{\Lambda}) (1-a X) \right) + \left( 1 - (1 -a \sqrt{\Lambda}) (1-a X) \right)
\]

\[
= e^{-2 \sqrt{\Lambda} T} \left( -1 + 2 a \sqrt{\Lambda} + 1-a \sqrt{\Lambda} -a X + a^2 \sqrt{\Lambda} X \right) + \left( 1 - 1 + a \sqrt{\Lambda} + a X - a^2 \sqrt{\Lambda} X \right)
\]

\[
= e^{-2 \sqrt{\Lambda} T} \left( a \sqrt{\Lambda} - a X + a^2 \sqrt{\Lambda} X \right) + \left( a \sqrt{\Lambda} + a X - a^2 \sqrt{\Lambda} X \right).
\]
Dropping the $a^2$ terms we have:

\begin{equation}
B = a \left[ \left( \sqrt{\Lambda} + X \right) + e^{-2 \sqrt{\Lambda} T} \left( \sqrt{\Lambda} - X \right) \right].
\end{equation}

Through similar manipulations, we find:

\begin{equation}
C = a^2 \left[ \left( \sqrt{\Lambda} + X \right) \left( \sqrt{\Lambda} + Y \right) + e^{-2 \sqrt{\Lambda} T} \left( \sqrt{\Lambda} - X \right) \left( \sqrt{\Lambda} - Y \right) \right].
\end{equation}

Putting A, B and C together we have:

\[
\begin{split}
&G(x,y;g;t) \\ &= \frac{1}{a} \frac{4 \Lambda e^{-2 \sqrt{\Lambda} T} }{\left[ \left( \sqrt{\Lambda} + X \right) + e^{-2 \sqrt{\Lambda} T} \left( \sqrt{\Lambda} - X \right) \right] \left[ \left( \sqrt{\Lambda} + X \right) \left( \sqrt{\Lambda} + Y \right) + e^{-2 \sqrt{\Lambda} T} \left( \sqrt{\Lambda} - X \right) \left( \sqrt{\Lambda} - Y \right) \right]}.
\end{split}
\]
Thus, using (4.2.4) with $\eta=1$:

\[
G_{\Lambda}(X,Y;T) = \lim_{a\to 0} a \, G(x,y;g;t)
\]

\begin{equation}
\begin{split}
\Rightarrow & G_{\Lambda}(X,Y;T) \\ &= \frac{4 \Lambda e^{-2 \sqrt{\Lambda} T} }{\left[ \left( \sqrt{\Lambda} + X \right) + e^{-2 \sqrt{\Lambda} T} \left( \sqrt{\Lambda} - X \right) \right] \left[ \left( \sqrt{\Lambda} + X \right) \left( \sqrt{\Lambda} + Y \right) + e^{-2 \sqrt{\Lambda} T} \left( \sqrt{\Lambda} - X \right) \left( \sqrt{\Lambda} - Y \right) \right]}.
\end{split}
\end{equation}

At this point several limits are contemplated (some of which will be used later on), and the method of obtaining $G_{\Lambda}(L_{1},L_{2};T)$ forms is discussed.

$G_{\Lambda}(L_{1},L_{2};T)$ can be obtained by performing an inverse Laplace transformation on $G_{\Lambda}(X,Y;T)$. Alternatively, one can take the $a \to 0$ limit of $G_{\lambda}(l_1,l_2;t)$ as defined in terms of a discrete inverse Laplace transformation in the last section. Either way we end up with the form:

\begin{equation}
G_{\Lambda}(L_{1},L_{2};T) = \int_{-i \infty}^{i \infty} \frac{d X}{2 \pi i} \int_{-i \infty}^{i \infty} \frac{d Y}{2 \pi i} e^{X L_1} e^{Y L_2} G_{\Lambda}(X,Y;T)
\end{equation}

Consider now the $T \to \infty$ limit (this and the $T \to 0$ limit will be useful for later derivations). Applying this limit to (4.2.18), we find:

\begin{equation}
G_{\Lambda}(X,Y;T) \xrightarrow{T \to \infty} \frac{4 \Lambda e^{-2 \sqrt{\Lambda} T}}{\left( X + \sqrt{\Lambda} \right)^2 \left( Y + \sqrt{\Lambda} \right)}
\end{equation}
An inverse Laplace transformation of this gives:

\begin{equation}
G_{\Lambda}(L_1,L_2;T) \xrightarrow{T \to \infty} 4 L_1 e^{-\sqrt{\Lambda} (L_1+L_2)} e^{-2 \sqrt{\Lambda} T}
\end{equation}

In the $T \to 0$ limit, applied to (4.2.18), we obtain:

\begin{equation}
G_{\Lambda}(X,Y;T) \xrightarrow{T \to 0} \frac{1}{X+Y}
\end{equation}

To examine the $T \to 0$ limit for $G_{\Lambda}(L_{1},L_{2};T)$, plug (4.2.22) into (4.2.19):

\[
G_{\Lambda}(L_{1},L_{2};T) \xrightarrow{T \to 0} \int_{-i \infty}^{i \infty} \frac{d X}{2 \pi i} \int_{-i \infty}^{i \infty} \frac{d Y}{2 \pi i} e^{X L_1} e^{Y L_2} \frac{1}{X+Y}
\]
For clarity, apply the substitution $X=i X'$, $Y=i Y'$.

\[
\Rightarrow G_{\Lambda}(L_{1},L_{2};T) \xrightarrow{T \to 0} \int_{\infty}^{-\infty} \frac{i d X'}{2 \pi i} \int_{\infty}^{-\infty} \frac{i d Y'}{2 \pi i} e^{i X' L_1} e^{i Y' L_2} \frac{1}{i(X'+Y')}
\]

\begin{equation}
\Rightarrow G_{\Lambda}(L_{1},L_{2};T) \xrightarrow{T \to 0} -i \int_{-\infty}^{\infty} \frac{d X'}{2 \pi} \int_{-\infty}^{\infty} \frac{d Y'}{2 \pi} e^{i X' L_1} e^{i Y' L_2} \frac{1}{X'+Y'}
\end{equation}
Doing the $Y'$-integral first:

\[
\int_{-\infty}^{\infty} \frac{d Y'}{2 \pi} e^{i Y' L_2} \frac{1}{X'+Y'} = \frac{e^{-i X' L_2}}{2 \pi} \int_{-\infty}^{\infty} d Y' \frac{e^{i L_2 (X'+Y')}}{X'+Y'}
\]

One may write the integral in the form of the exponential integral $Ei(i L (X'+Y'))$, which is defined by:

\begin{equation}
Ei(x) = - \int_{-x}^{\infty} dt \frac{e^{-t}}{t}
\end{equation}
which can be written in terms of the E$n$-function with $n=1$:

\begin{equation}
E_1(x) \equiv \int_{1}^{\infty} dt \frac{e^{-t x}}{t} = \int_{x}^{\infty} du \frac{e^{-u}}{u},
\end{equation}
so that:

\begin{equation}
E_1(x) = -Ei(-x)
\end{equation}

To see this, consider:

\[
\frac{d}{d Y'} Ei \left( i L (X'+Y') \right) = \frac{d}{d Y'} \left( -E_1 \left( - i L (X'+Y') \right) \right) = - \frac{d}{d Y'} \int_{1}^{\infty} dt \frac{e^{i t L (X'+Y')}}{t}
\]

\[
= - \int_{1}^{\infty} dt (i t L) \frac{e^{i t L (X'+Y')}}{t}
= - i L \int_{1}^{\infty} dt e^{i t L (X'+Y')}
\]

\[
= - i L \left[ \frac{e^{i t L (X'+Y')}}{i L (X'+Y')} \right]_{t=1}^{t=\infty}
= \frac{e^{i L (X'+Y')}}{X'+Y'}
\]
Thus:

\[
\frac{e^{-i X' L_2}}{2 \pi} \int_{-\infty}^{\infty} d Y' \frac{e^{i L_2 (X'+Y')}}{X'+Y'} = \frac{e^{-i X' L_2}}{2 \pi} \int_{-\infty}^{\infty} d Y' \left( \frac{d}{d Y'} Ei \left( i L_2 (X'+Y') \right) \right)
\]

\[
\Rightarrow \int_{-\infty}^{\infty} \frac{d Y'}{2 \pi} e^{i Y' L_2} \frac{1}{X'+Y'} = \frac{e^{-i X' L_2}}{2 \pi} \left[ Ei \left( i L_2 (X'+Y') \right) \right]_{-\infty}^{\infty}
\]
Now,

\begin{equation}
\lim_{x \to \infty} Ei(i x) \to i \pi, \quad \lim_{x \to - \infty} Ei(i x) \to - i \pi
\end{equation}

\[
\Rightarrow \int_{-\infty}^{\infty} \frac{d Y'}{2 \pi} e^{i Y' L_2} \frac{1}{X'+Y'} = \frac{e^{-i X' L_2}}{2 \pi} \left[ 2 \pi i \right]
\]

\begin{equation}
\Rightarrow \int_{-\infty}^{\infty} \frac{d Y'}{2 \pi} e^{i Y' L_2} \frac{1}{X'+Y'} = i e^{-i X' L_2}
\end{equation}

Plugging this result into (4.2.23):

\[
G_{\Lambda}(L_{1},L_{2};T) \xrightarrow{T \to 0} -i \int_{-\infty}^{\infty} \frac{d X'}{2 \pi} e^{i X' L_1} i e^{-i X' L_2}
\]

\[
= \int_{-\infty}^{\infty} \frac{d X'}{2 \pi} e^{i X' (L_1-L_2)}
\]

\begin{equation}
\Rightarrow G_{\Lambda}(L_{1},L_{2};T) \xrightarrow{T \to 0} \delta(L_1-L_2)
\end{equation}

This is just as we would expect: the probability of propagating from a loop of length $L_1$ to a loop of length $L_2$ in the limit where the time for the propagation $T \to 0$ should be zero unless $L_1=L_2$, which is to say that the loop length remains unchanged.

The authors give the final formula for $G_{\Lambda}(L_{1},L_{2};T)$ in the general case, obtained by performing an inverse Laplace transform of (4.2.18):

\begin{equation}
G_{\Lambda}(L_{1},L_{2};T) = \frac{e^{-[\coth{\Lambda T}] \sqrt{\Lambda} (L_1 + L_2)}}{\sinh{\sqrt{\Lambda} T}} \frac{\sqrt{\Lambda L_1 L_2}}{L_2} I_1 \left( \frac{2 \sqrt{\Lambda L_1 L_2}}{\sinh{\Lambda T}} \right),
\end{equation}
where $I_1(x)$ is a modified Bessel function of the first kind.

One may compute the expression for the probability of propagating from $L_1$ to $L_2$ for an arbitrary step length $T$ by integrating the above expression over $T$ (from $0 \to \infty$). Thus we obtain:

\begin{equation}
G_{\Lambda}(L_{1},L_{2}) = \int_{0}^{\infty} d T G_{\Lambda}(L_{1},L_{2};T) = \frac{e^{-\sqrt{\Lambda} | L_1 - L_2 |}-e^{-\sqrt{\Lambda} (L_1 + L_2)}}{2 \sqrt{\Lambda} L_2}.
\end{equation}

One final point that must be made about the continuum limit involves the analytic continuation of the space-time variables. As has been mentioned, obtaining the path integrals for the Lorentzian theory is a matter of performing the substitution $\Lambda \to -i \Lambda$. One is naturally led to consider the analytical continuation of what we might have considered ``time": our variable $T$. Attempting to use $T \to -i T$ in (4.2.30), for example, yields very different (and singular) results. As the authors reason, however, this choice for the analytic continuation of $T$ is wrong.

To understand the proper way to consider analytic continuation in time, one has to trace the origin of the $T$-terms in the continuum limit equations. The term $T$ always appears in the combination $\sqrt{\Lambda} T$ , which originated by taking the continuum limit of $F^t$-like terms in the discrete expressions of the previous section.

Thus we consider instead the analytic continuation of $F^t$. The $t$ term is not to be analytically continued since it is merely a counter for the amount of iterations performed. It is $F$ itself that is to be analytically continued, since its definition in terms of $g$ (seen earlier) relate it to the action. To see this, first recall the definition of $g$:

\begin{equation}
g = e^{i \lambda a^2} = e^{i \lambda a_t a_l},
\end{equation}
where we have distinguished $a_t$, the time-direction lattice spacing, from $a_l$, the space-direction spacing, for clarity.

Looking back at (4.2.11), this means:

\[
g=\frac{1}{2} e^{-\Lambda a_t a_l} \to \frac{1}{2} \left( 1 - \frac{1}{2} \Lambda a_t a_l \right), \quad ({\rm i.e.} \; F=1-\sqrt{a_t a_l \Lambda}=e^{-\sqrt{a_t a_l \Lambda}}).
\]

From this definition for F we can see that its analytic continuation in time would mean taking $a_t \to -i a_t$; that is, the transformation involves converting the length of the time-like spacing from Euclidean to Lorentzian. This gives exactly the same result as instead applying the analytic continuation to the cosmological constant, $\Lambda \to -i \Lambda$. Thus we see that thinking of $T$ as a physical ``time"-parameter is a mistake. Consequentially, as the authors point out, a Hamiltonian derived using it would also be physically irrelevant.

\subsection{The Differential Equation, Disk Amplitudes and more}

The aim of this section is to address a few other results discussed in \cite{ambjorn98}. An interesting point is that one can obtain (4.2.18) by taking the continuum limit of the recursion relation (4.1.16). Inserting the relations (4.2.11) and (4.2.12) into (4.1.16) and expanding to first order in the lattice spacing one can obtain:

\begin{equation}
\frac{\partial}{\partial T} G_{\Lambda}(X,Y;T) + \frac{\partial}{\partial X} \left[ (X^2 - \Lambda) G_{\Lambda}(X,Y;T) \right] = 0.
\end{equation}
The PDE is solved using (4.2.20) as a boundary condition at $T=0$. The solution is:

\begin{equation}
G_{\Lambda}(X,Y;T) = \frac{\bar{X}^2(T;X)-\Lambda}{X^2 - \Lambda} \frac{1}{\bar{X}(T;X)+Y},
\end{equation}
where $\bar{X}(T;X)$ is the solution to the characteristic equation:

\begin{equation}
\frac{d \bar{X}}{d T} = -(\bar{X}^2-\Lambda), \quad \bar{X}(T=0)=X.
\end{equation}

Solving this relation and plugging into (4.3.2) indeed results in (4.2.18). The authors then use this differential equation to obtain the Hamiltonian for the system and construct the solution to the ``Wheeler-DeWitt equation". However, as mentioned in the article and at the end of the previous section, the Hamiltonian obtained in this manner cannot be considered physically relevant - it corresponds to the parameter $T$ which does not correspond to physical time for the system.

The next point to be made involves the so-called ``disc amplitude", also called the Hartle-Hawking wave function, for the system. The disc amplitude gives the probability of a spatial slice of loop length L collapsing to a loop of zero length at arbitrary time (alternatively, the probability of creation of a spatial slice of loop length L from nothing at arbitrary time):

\begin{equation}
W_{\Lambda}(L) \equiv G_{\Lambda}(L, L_2=0).
\end{equation}

We can find the exact expression for the disc amplitude in our model by applying the $L_2 \to 0$ limit to (4.2.31). In this limit $L_1 > L_2$, therefore:

\[
\begin{split}
\frac{e^{-\sqrt{\Lambda} | L_1 - L_2 |}-e^{-\sqrt{\Lambda} (L_1 + L_2)}}{2 \sqrt{\Lambda} L_2} & \xrightarrow{L_1 > L_2} e^{-\sqrt{\Lambda} L_1} \frac{e^{\sqrt{\Lambda} L_2}-e^{-\sqrt{\Lambda} L_2}}{2 \sqrt{\Lambda} L_2} 
= e^{-\sqrt{\Lambda} L_1} \frac{\sinh \sqrt{\Lambda} L_2}{\sqrt{\Lambda} L_2} \\
& \xrightarrow{L_2 \to 0} e^{-\sqrt{\Lambda} L_1}
\end{split}
\]

\begin{equation}
\Rightarrow W_{\Lambda}(L) = e^{-\sqrt{\Lambda} L}.
\end{equation}

The disc amplitude is then used in the source to compare to the one resulting from the older Euclidean model. The details of the comparison seem to draw on Dynamical Triangulations results too heavily for the purposes of this paper; the main point is that they differ.

One more interesting result that can be drawn from our continuum path integral and which can be compared to the corresponding Euclidean result is the average spatial volume $\langle L_{space} \rangle$.

In the Euclidean model, the following relation was obtained:

\begin{equation}
G_{\Lambda}^{(eu)}(L_1,L_2;T) \propto e^{-\sqrt[4]{\Lambda} T} \quad for \quad T \to 0.
\end{equation}
We can use this to compute the average \emph{two-dimensional} volume $V(T)$:

\[
\begin{split}
\langle V(T) \rangle^{(eu)} &= - \frac{1}{G_{\Lambda}^{(eu)}(L_1,L_2;T)} \frac{\partial}{\partial \Lambda} 
G_{\Lambda}^{(eu)}(L_1,L_2;T) \\
& \propto - \frac{1}{e^{-\sqrt[4]{\Lambda} T}} \left( - \frac{1}{4} \Lambda^{-3/4} T \right) e^{-\sqrt[4]{\Lambda} T}
\end{split}
\]

\begin{equation}
\Rightarrow \langle V(T) \rangle^{(eu)} \propto \frac{T}{\Lambda^{3/4}}.
\end{equation}
For large $T$ we expect the average spatial volume at intermediate T's to behave like:

\begin{equation}
\langle L_{space} \rangle^{(eu)} = \frac{\langle V(T) \rangle^{(eu)}}{T} \propto \frac{1}{\Lambda^{3/4}}.
\end{equation}

Compare now to our own model. From (4.2.21):

\begin{equation}
G_{\Lambda}(L_1,L_2;T) \propto e^{-\sqrt{\Lambda} T}.
\end{equation}
Applying the same methods as we just did for the Euclidean model, we find:

\begin{equation}
\langle V(T) \rangle \propto \frac{T}{\sqrt{\Lambda}}.
\end{equation}

\begin{equation}
\langle L_{space} \rangle \propto \frac{1}{\sqrt{\Lambda}}.
\end{equation}

This reflects the fact that the CDT quantum space-time we are working with does not have the anomalous fractal dimension that characterized two-dimensional Euclidean quantum gravity: $\sqrt{\Lambda}$ has dimension $1/[L]$, and thus our average spatial volume (which is of course one-dimensional in our model) has the unsurprising dimension $[L]$.

\newpage

\subsection{Topology}

The purpose of this final section is to mention some of the work done by the CDT authors in considering topology in the context of CDT.

In \cite{ambjorn98}, topology change is addressed in the context of allowing the spatial topology to change as a function of time. This means that a ``baby universe" is allowed to branch off from the main one at some time $T$, eventually disappearing into the vacuum - it is not allowed to rejoin the ``parent" universe. This restriction (and, indeed, the entire consideration of this form of topology change) is imposed to permit a comparison with the analogous calculation in previous 2-D Euclidean calculations. The actual process is taken to be forbidden in CDT (at the time the paper was published) due to causality violations; however, some discussion of ways of possibly allowing this process in some form under the CDT model has been under way for some time and will be mentioned later on. The details of the comparison with DT and the meaning of many of the mathematical objects derived to allow for said comparison rely too heavily on knowledge of prior Euclidean work to be of relevance to this report; however, a qualitative consideration of the basic idea of how to represent the topology change seems worthwhile.

\begin{figure}[ht]
\begin{minipage}[b]{0.5\linewidth}
\centering
\includegraphics[scale=0.5]{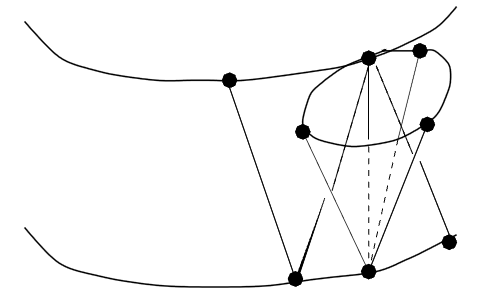}
\caption{A “baby universe” branches off locally in one time-step. \cite{ambjorn98}}
\end{minipage}
\hspace{0.5cm}
\begin{minipage}[b]{0.5\linewidth}
\centering
\includegraphics[scale=0.5]{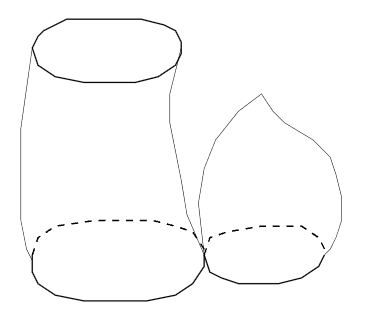}
\caption{A “baby universe” is created by a global pinching of the spatial loop. \cite{ambjorn98}}
\end{minipage}
\end{figure}

The process of the branching off of a ``baby universe" is shown in Figure 5. Figure 6 shows an alternative and technically simpler way to represent the process, and the derivation follows that representation (the continuum limit is the same in both cases).

The modification of our approach begins with the 1-step discrete propagator, which takes the form:

\begin{equation}
G_{\lambda}(l_1,l_2;1) = G_{\lambda}^{(b)}(l_1,l_2;1) + \sum_{l=1}^{l_1-1} l_1 w(l_1-l,g) G_{\lambda}^{(b)}(l,l_2;1),
\end{equation}
where $G_{\lambda}^{(b)}$ refers to the ``bare" propagator without topology changes. The added term consists of the path integral for propagating from a spatial slice of length $l$ to one of length $l_2$ (in one step), multiplied by the discrete disc amplitude corresponding to a loop of length $l_1 - l$ (the length left over by the propagator). This is multiplied by $l_1$, representing the fact that the ``pinching" may occur at any of the $l_1$ vertices. Finally this term is summed over all possible ways of separating the full length $l_1$ into two. The discrete disc amplitude $w(l,g)$ (the exact nature of which is about to be discussed) gives the amplitude that a loop of length $l$ collapses to a loop of one vertex (in the continuum limit this becomes a loop of zero length) in arbitrary time.

The nature of $w(l,g)$ in this formula is not straightforward; it is not the ``bare" disc amplitude, but rather some modified version of it that accounts for topology changes (its exact form is unclear and beyond the scope of this section, depending heavily on analogues with the earlier Euclidean models). The bare disc amplitude is the discrete version of (4.3.5), and can be easily constructed by analogy with (4.3.4) (and (4.2.31) in turn for the definition of $G_{\Lambda}(L_1,L_2)$):

\begin{equation}
w^{(b)}(l,g) \equiv \sum_t G^{(b)}(l,l_2=1;g;t) = G^{(b)}(l,l_2=1;g).
\end{equation}

Note that, as suggested above, the second argument of $G^{(b)}$ is $l_2=1$ rather than 0, which was the argument in the continuum version of these equations; this is because $l=0$ is a singular case in the discrete version, which was in fact removed manually from the path integral in the discussion preceding (4.1.13). The contact with the continuum limit argument $L_2=0$ becomes evident when one considers the relation $L_2=l_2 a$, with $l_2=1$ and $a \to 0$.

The purpose of the rest of \cite{ambjorn98} is to find an expression for $W_{\Lambda}(X)$ under this ``baby universe" regime and compare with the same result from 2-D Euclidean quantum gravity; this is accomplished and they are found to match after rescaling.

Another aspect of topology and CDT involves the inclusion of a sum over topologies in the path integral. In \cite{ambjorn06}, it is argued that a sum over topologies would cause the path integral too diverge too badly to be included in the expression for the path integral. \cite{loll06} and other papers attempt to find a way to circumvent this problem. This work is too involved for the scope of this report; the main premise is that the problem may be solvable by imposing some sort of causal restrictions to which geometries the sum over topologies can add to the path integral.

\newpage

\section{Conclusion}

In this report, the basic ideas of CDT have been introduced and its application to two-dimensional spacetimes presented. While two-dimensional geometry considerations are very different from those for three or four dimensions, the two-dimensional case is useful for illustrating concepts one cannot easily visualize in higher dimensional models. An additional reason for studying the two-dimensional case is that much of current CDT work (in four dimensions) relies on heavy computer simulation, but the two-dimensional case can be solved analytically, with its expressions and assumptions more easily identifiable and verifiable. The details of accounting for topology changes are unfortunately beyond the scope of the report, but an understanding of the process in two dimensions is crucial to attempting to account for it in three or more dimensions.

As mentioned in the introduction, one reason this topic was chosen for this research project is curiosity regarding its relative obscurity.
Some of the possible reasons for CDT's apparent lack of popularity and low number of contributing authors may be its relative youth and a certain lack of marketing: there is a lack of popular and less mathematically intensive sources on the subject. The hope is that this report offers a guide to the basics of CDT and how to derive many of the expressions its 2-D application, and is helpful in providing a starting point from which to investigate CDT in more depth.

\end{document}